# Inclusion in Virtual Reality Technology: A Scoping Review

Xiaofeng Yong[1*], Ali Arya[1]

[1]School of Information Technology, Carleton University, Ottawa, Ontario, Canada

**Abstract**

Despite the significant growth in virtual reality applications and research, the notion of inclusion in virtual reality is not well studied. Inclusion refers to the active involvement of different groups of people in the adoption, use, design, and development of VR technology and applications. In this review, we provide a scoping analysis of existing virtual reality research literature about inclusion. We categorize the literature based on target group into ability, gender, and age, followed by those that study community-based design of VR experiences. In the latter group, we focus mainly on Indigenous Peoples as a clearer and more important example. We also briefly review the approaches to model and consider the role of users in technology adoption and design as a background for inclusion studies. We identify a series of generic barriers and research gaps and some specific ones for each group, resulting in suggested directions for future research.

Keywords: Inclusion, Virtual Reality, Literature Study, Research Trends

## 1    Introduction

Technological advances are driven by a variety of factors, such as scientific curiosity (Nowotny, 2010), economic development (Nicholas, 2011; Taalbi, 2017), and social needs (ESDC, 2018; Wilburn & Wilburn, 2014). The prioritization of economic growth and commercial profit in private sector, has called for changes in government funding and regulations (ESDC, 2018) and business models (Wilburn & Wilburn, 2014) to support technical innovation for social and environmental benefits. The interests, requirements, and concerns of different groups of people have always influenced technical developments, but in many cases, they have been overshadowed by mainstream demands and commercial considerations (Daniels & Geiger, 2010; Mott, et al., 2019), frequently considering young males, high-end professionals, or other specific groups (Sieß, Beuck, & Wölfel, 2017). This is particularly evident and problematic for marginalized groups such as Indigenous Peoples and other racialized communities (ethnic minorities), women, and LGBTQ+. Similarly, age and mental/physical abilities are factors that can have a significant effect on technology use but may not be considered properly by technology designers. Emerging technologies, such as the Internet of Things (IoT), Artificial Intelligence (AI), and Virtual Reality (VR), can be essential in serving many users. However, the diversity of user groups and their specific concerns have frequently been an afterthought of technology design if considered at all, and the underrepresentation of many marginalized groups has resulted in design processes and, consequently, products that are not inclusive. For example, age and gender are suggested to have effects on the cybersickness (Petri, Feuerstein, Folster, Bariszlovich, & Witte, 2020) or the sense of presence (Lorenz, Brade, Klimant, Heyde, & Hammer, 2023) in VR applications. As such, solutions for cybersickness or increased presence need to take these factors into account, so they can be inclusive and cater to a wider range of users. By inclusion, we refer to not only reflecting the needs, opinions, interests, traditions, and protocols of all groups but also allowing their active participation and making sure they own and benefit from the designed products and services (NSERC, 2017). While VR is shown to be an effective educational tool (Scavarelli, Arya, & Teather, 2021), its use for some groups, such as children, is limited due to these age and gender effects and also lack proper educational content (Mado, et al., 2022). As another example, an immersive storytelling experience that is not designed with the participation of Indigenous people is not likely to respect their unique ownership protocols and cultural sensitivities and, as such, will not be authentic or beneficial for them or any other user group (Winschiers-Theophilus, 2022; Tubby, 2022). The lack of inclusion means that technology is not properly beneficial to many users and also that technology providers are missing a potentially significant user base or forcing them to use products that are not suitable for their lived experience (Reid & Gibert, 2022).

With the introduction of affordable VR devices, such as Oculus Rift and its development kit, an increased number of research projects related to VR have appeared in the literature. However, as Alcañiz et al. (2019), and Steffen et al. (2019) pointed out that the knowledge of this field is significantly fragmented and focused largely on specific domains. Despite its transformational potential to change the way we interact with the information (Steffen, Gaskin, Meservy, Jenkins, & Wolman, 2019), common (particularly, commercial) VR systems and experiences have only recently started to pay attention to inclusive concerns (Dombrowski, Smith, Manero, & Sparkman, 2019; Menke, Beckmann, & Weber, 2019). Interaction and device accessibility, inclusive representation, and application diversity are among topics that have been frequently ignored or considered as an afterthought (Mott, et al., 2019). While there is a trend toward increasing the accessibility of VR for people with physical disabilities (Mott, et al., 2019), there is exceedingly little research on the ethical, cultural, and usability protocols and perspectives to be considered when designing VR experiences appropriate for marginalized communities, such as LGBTQ+ and Indigenous Peoples (Yong, Arya, & Manatch, 2023). These protocols can deal with concerns such as individuality, ownership, authenticity, presence, embodiment, and ergonomics.

Recent studies suggest the presence of gender-based differences among individuals experiencing VR (Felnhofer, Kothgassner, Beutl, Hlavacs, & Kryspin-Exner, 2012; Munafo, Diedrick, & Stoffregen, 2017). The resulting issues range from hardware fitness for female users (Menshikova, Tikhomandritskaya, Saveleva, & Popova, 2018) to social concerns such as "safe spaces" for LGBTQ+ (Acena & Freeman, 2021). Other examples include issues such as body representation, forms of interaction, privacy, and intimacy (Maloney, Freeman, & Robb, 2021). The research on gender bias has had some initial results regarding differences between the experiences of male and female VR users, but it has not been comprehensive and conclusive (Grassini & Laumann, 2020). Few studies have started to investigate the experience of LGBTQ+ VR users (Acena & Freeman, 2021; Paré, Sengupta, Windsor, Craig, & Thompson, 2019), and Indigenous, black, and other ethnic communities can have established knowledge, plans, and protocols that go beyond individual concerns. While an individual user may be happy with the mere ability to use a VR system, a community is a holistic entity with its own tradition, protocols, and knowledge (Cattaneo, Giorgi, Herrera, & del Socorro Aceves Tarango, 2022; Guerrero Millan, 2023) and may have concerns such as appropriate use of cultural artifacts and group ownership of created assets. The interaction of Indigenous Peoples and ethnic communities with VR technology involves many unanswered questions and unresolved problems, including how to design systems that allow these communities to collectively own and control their data, incorporate their traditional knowledge, present their stories and opinions, and benefit from the results in line with their collective plans. Recently-developed frameworks for collaborating with Indigenous Peoples and managing the resulting data such as the

Ownership, Control, Access, and Possession (FNIGC, 2014), Findable, Accessible, Interoperable, and Reusable (FAIR) (GO-FAIR, n.d.), and Collective Benefit, Authority to control, Responsibility, and Ethics (CARE) (GIDA, 2019) are starting to appear in data-heavy subjects such as AI (Lewis, et al., 2020) but have not been considered within the context of interactive technologies such as VR.

To address the issue of user inclusion, methodologies such as User-Centred Design (UCD) and Participatory Design (PD) (Abras, Maloney-Krichmar, Preece, & others, 2004) promote a design process that engages users directly. However, while they have been successful in establishing a connection between technology developers and users, they do not provide specific guidelines for working with communities, as opposed to a series of individual users (Fleury & Chaniaud, 2023). The target groups are commonly defined statistically as those with similar demographics but are not necessarily represented as holistic entities with group values and characteristics. A paradigm shift for looking at users not as a group of individuals but as members of a holistic community with its own characteristics have been pursued in areas such as community psychology, and researchers have suggested generic approaches for co-production with communities that include interdisciplinarity, public value, authenticity, and reflection (Preece & Maloney-Krichmar, 2003; Bergvall-Kareborn & Stahlbrost, 2009; O'Donnell, et al., 2016; Cattaneo, Giorgi, Herrera, & del Socorro Aceves Tarango, 2022; Guerrero Millan, 2023). Despite such efforts, incorporating users in the design process as communities with group knowledge, protocols, beliefs, and traditions through well-defined practices has not been properly explored within the context of technology design and usage models, especially for emerging technologies such as VR. Common initiatives to study and represent technology usages, such as the Technology Acceptance Model (TAM) (Abrahamson, 1991) and Diffusion of Innovations (DOI) (Rogers, Diffusion of Innovations, 2010) are mostly about individuals (or sets of individuals (Fleury & Chaniaud, 2023)), households, and occasionally organizations as opposed to larger communities with a long history, traditions, and protocols (O'Donnell, et al., 2016).

Despite the significant growth in VR research and development and a wide range of applications, the research on the inclusion of VR technology is rather limited. In this paper, we aim at a critical review of the literature on this subject. Our specific research questions are:

(1) What barriers and solutions for inclusion in VR have been investigated?
(2) Which areas and aspects of inclusion in VR are not properly explored?
(3) What are potential solutions and research directions to improve VR inclusion?

Potential VR users form a very wide range of groups, categorized based on many characteristics. These personal and group characteristics can form different barriers to inclusion. As discussed in our review, the existing research has focused mainly on ability, gender, and age as the basis of individual diversity. These topics emerged among the primary inclusive concerns in the reviewed VR literature. We also noticed and reviewed a relatively new trend distinguishing between individuals and communities with approaches to community-based design in VR. We define the term community as "a body of persons or nations having a common history or common social, economic, and political interests" (Merriam-Webster, 2023). Ethnic groups, such as Indigenous Peoples, are the most common example in the community-based papers we have reviewed. The research papers about working with groups of people with common age or gender characteristics are not categorized as community-based in our review, even though they may have used the term and their target groups for under other definitions of community. The rationale for this decision is that the defining characteristics of these groups have been mainly individual, as opposed to social features such as culture and tradition in, e.g., ethnic groups. The term Indigenous, in this context, refers to "the earliest known inhabitants of a place and especially of a place that was colonized by a now-dominant group" (Merriam-Webster, Indigenous definition, 2021). Our review does not aim to show the diversity of VR applications in terms of purpose and field of work unless the application directly relates to including certain user groups within the scope of our review.

This review contributes to the field of virtual reality by identifying (1) the current ways that inclusion is incorporated and investigated, (2) the knowledge gaps and open problems, and (3) suggestions for further research. It does not intend to show how much of VR research is dealing with inclusion and which VR initiatives are not addressing inclusion but can identify the inclusion trends and be the basis for new research projects and future, more targeted, systematic reviews on specific aspects of inclusion in VR. The structure of the review follows the PRISMA extension for scoping reviews (PRISMA, n.d.). Following this introduction, we describe the literature search and analysis method, briefly review the main concepts in technology design and adoption as the necessary basis for understanding inclusion, and then present and discuss the review of significant literature on inclusion in VR.

## 2 Method

The notion of inclusion is two-fold and consists of including different groups as users and allowing them to participate in the design (and development) process. As such, we start our study (in Section 3) with a review of existing models and approaches for the design processes and the role of users in them, such as the Technology Acceptance Model (TAM), Diffusion of Innovation (DOI), and User-Centred Design (UCD). This part of our study is not meant to be exhaustive or a scoping review. It only reviews some of the most relevant approaches to consider users in the design process of technology products and systems. While independent of a particular technology, these approaches have been mentioned by the papers we review or in our discussion. Their brief overview will help us understand the next (and the main) part of our review, which is specifically about inclusion in VR.

To capture the full picture of the field of VR with respect to inclusion and our research questions, we chose the Scoping Review extension of the well-established Preferred Reporting Items for Systematic Reviews and Meta-Analysis Protocols (PRISMA-ScR) (PRISMA, n.d.). We chose to do a scoping review because the topic of inclusion in VR is a fairly new subject of study and there are not many literature reviews on this topic. Following the possible intentions listed by Munn et al. (Munn, Peters, & Stern, 2018), we intended to do this work as a first step to identify the types of related work, key concepts, and knowledge gaps as a precursor to more systematic reviews on specific aspects. As such, we were more interested in identifying trends than listing all works that are somewhat related. After Protocol selection, PRISMA-ScR steps includes eligibility criteria, information sources, search, selection, data charting and items, appraisal, and analysis, which we briefly discuss in this section.

We included papers that specifically addressed inclusion in VR and were written in English (as the primary language in VR research and our team). There is a large body of publications on VR with topics from rendering to interaction methods, but mostly without discussion of specific user or developer groups. As the result of excluding such work, our review does not answer questions such as how inclusion research is growing relative to other aspects of VR research. We only wanted to see what inclusion concerns and solutions have been discussed by VR researchers. We did not enforce any time range or source limitations, and included some non-VR sources that provided general information on inclusion and design methodologies relevant to our discussion.

We searched in Google Scholar that provides a general coverage of research publications and directly in the ACM Digital Library and IEEE Xplore databases, which together publish or index most of the technical research in VR. We conducted several rounds of literature search by employing a snowball method (Wohlin, 2014). To begin with, we used the keywords "Inclusive VR," "VR bias," "age AND VR," "gender AND VR," "accessibility AND VR," "ability AND VR," "community AND VR," "Indigenous AND VR", and "community AND User-Centred Design". From the papers on the first round of snowball approach, we found and added other keywords including "VR AND ethical issues," "VR AND empathy," "VR AND user acceptance," and "VR AND representation". A total of 48 out of 63 papers were selected at the initial phase as seed papers, covering a wide range of topics on inclusion in VR. We then performed forward and backward snowball and later re-did the search for more recent papers (2022 and 2023) to make sure no new development was missed after 2021, when our project started. We also included a few papers suggested by our external reviewers. Finally, we added some references for non-VR and more general sources related to technology adoption by different groups and user participation in the research and design process, as they were seen as relevant to and cited by VR research when inclusion is considered. Figure 1 shows the search process of literature search and selection.

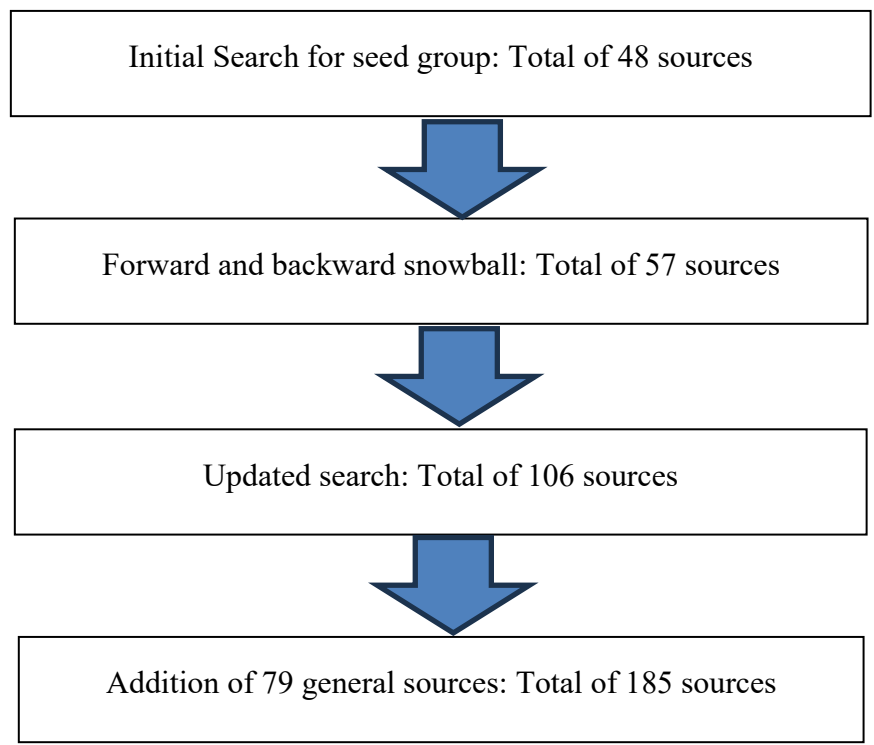

**Figure 1. Literature Search Process**

The total number of papers that were considered throughout the process was more than 500, but after excluding many (as described above), our selection process resulted in 185 sources. Out of that number, 43 were related to general topics such as the role of users in adopting and designing technology products, basic VR features, and cited research methods. Another 36 sources were related to inclusion but in a more general sense and not specific to VR. The remaining 106 sources were directly related to inclusion in VR. Figure 2 shows how the final sources are distributed by year. As mentioned, the numbers do not show all the publications related to inclusion, as we selected representative samples to understand the scope of inclusion research. However, the numbers do show the increasing attention to this topic.

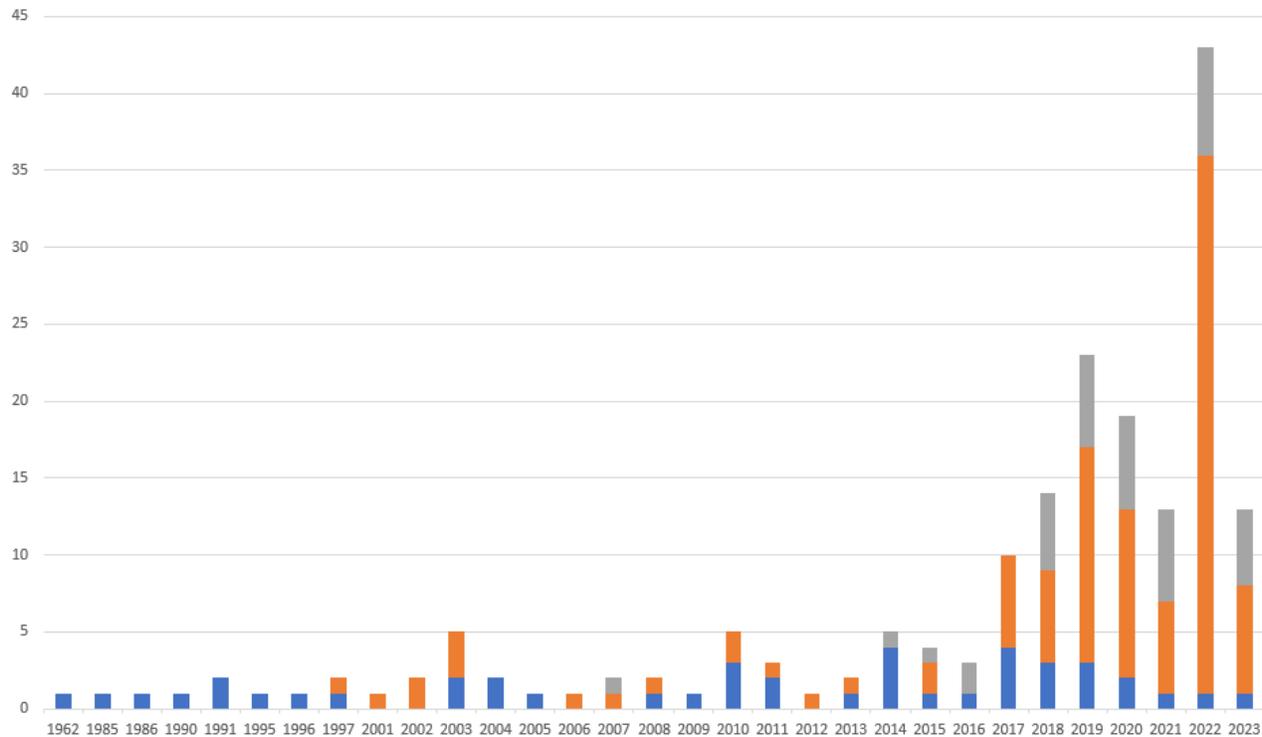

**Figure 2. Distribution of sources by year, for three groups: (A) General Sources, (B) General Inclusion, and (C) Inclusion in VR.**

**Sources from 2023 are from the first half of the year, prior to the submission of this paper.**

To review the selected research papers, we used a thematic analysis approach by coding the data as we looked for two types of themes: user groups and inclusion concerns. For each paper, we identified the specific user group(s) and inclusion concern(s) they had in addition to any proposed solution. We also recorded the year. As ours was not a full systematic review, we did not include more details although future studies could focus on specific areas of inclusion and then add details such as hardware and software technologies. We finally synthesized our results (Section 4) by categorizing the reviewed papers using user groups and then discussing the inclusive concerns for each. We also identified cross-cutting concerns and discussed them separately (Section 5.1).

## 3 Designing with Users

Literature on various inclusion aspects of VR directly or indirectly connects to the more general notion of designing technology products for different groups of users. As such, it is important to have a brief discussion of this notion to establish a base understanding. In this section, we provide a brief discussion, starting with theories on how user groups adopt a technology. We then discuss approaches to directly involve users in the design and finally review the issues when communities of users are considered, as opposed to individuals and the differences such a consideration causes.

### 3.1 Technology Adoption

The technology acceptance model (TAM) has gained popularity in explaining users' behaviour toward technology since its introduction in 1985 by Fred Davis (Davis, 1985). Rooted in the theory of reasonable action (TRA) and theory of planned behaviour (TPB), TAM has the mediating role of two variables, perceived ease of use and perceived usefulness between technology characteristics and potential system usage (Marangunić & Granić, 2015), as shown in Figure 3. Succeeding researchers (Abrahamson, 1991; Jiang, Chen, & Lai, 2010; Moore & Benbasat, 1991; Marangunić & Granić, 2015) have extended TAM to consider factors from related models, belief factors, personality traits, and demographic characteristics. Researchers have also suggested conducting studies on and developing models for communities (O'Donnell, et al., 2016) as TAM and other similar approaches are more suitable for individuals, households, and organizations (Koch, Toker, & Brulez, 2011) (O'Donnell, et al., 2016). Moreover, current TAM may not be able to measure and predict technology use across different cultures and gender because TAM emphasizes the individual's rational choice, highlighting the conscious willingness to make decisions of common sense, which may no longer be the case when certain technologies are deployed in a culture or in the contexts that are different from where they were designed and developed (Ajibade, 2018; Koch, Toker, & Brulez, 2011). Koch et al. (2011) emphasize the role of trust in this regard and state that technology acceptance and adoption depends not only on perceived usefulness and ease of use, but also on perceived community characteristics. They show that these characteristics go beyond perceived community size and include community structure and culture that is different from typical organizational studies on TAM.

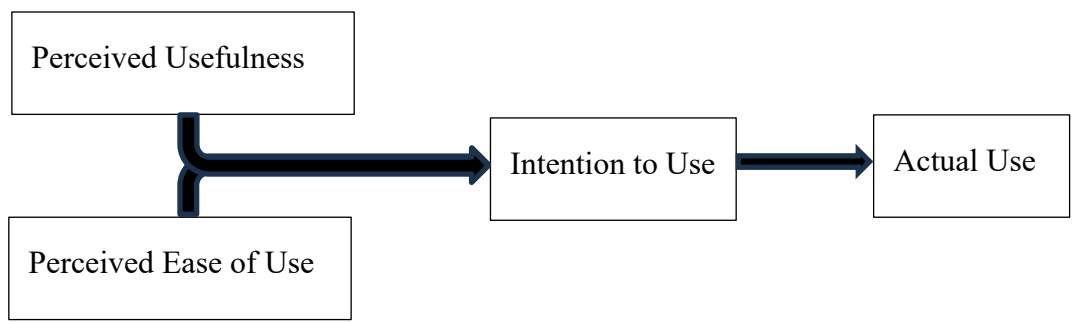

**Figure 3. Technology Acceptance Model**

Diffusion of Innovation (DOI) theory is another popular framework that has been used in many studies across many fields to understand the adoption of a given innovation (Rogers, 2010; Meyer, 2004; Richardson, Lingat, Hollis, & Pritchard, 2020). DOI has played an important role in the recent adoption of online learning technologies, such as using the theory framework to identify how the DOI factors may influence the faculty members to transit from classroom-based teaching to an online or hybrid teaching environment (Bennet & Bennett, 2003) (Richardson, Lingat, Hollis, & Pritchard, 2020). While TAM and DOI have different names and introduced few decades apart, researchers have found the overlapping of the core constructs of the two frameworks, which are ease of use (TAM) vs. technical complexity (DOI), perceived usefulness (TAM) vs. relative advantage (DOI) (Bradford & Florin, 2003; Taylor & Todd, 1995; Crum, Premkumar, & Ramamurthy, 1996; Tornatzky & Fleischer, 1990). However, DOI also contains Time as a factor in the framework, as DOI has focused on the diffusion rates and the sequence of when adoption occurs in a longitudinal setting (Rogers, 2010). Furthermore, Richardson et. al. (2020) have expanded the diffusion of innovation theory framework where they have combined Rogers' original DOI framework (Rogers,

1962), with relative advantage, compatibility, complexity, trainability, and observability, while Moore and Benbasat (1991) have added image, voluntariness and demonstrability.

Overall and despite their strengths, both TAM and DOI rely primarily on the individual generic users and lack consideration for communities with their own collective knowledge and tradition and different personal characteristics of users, including community, culture, age, gender, abilities, and social background, are not directly considered in these models (Rogers, 2010; Ajibade, 2018; Koch, Toker, & Brulez, 2011; O'Donnell, et al., 2016; Abrahamson, 1991).

### 3.2 User-based Design

User-Centred Design (UCD) is a broad term referring to various design practices that allow users to influence the design of products and services (Abras, Maloney-Krichmar, Preece, & others, 2004). While the term was used as early as 1977, it was popularized by Donald Norman (Norman, 1986). UCD follows recommendations such as clear action options, noticeable feedback, ease of evaluation, and natural and intuitive behaviour of the designed products (Abras, Maloney-Krichmar, Preece, & others, 2004). The focus on usability and user considerations is even more important in light of many studies on various biases in technology design, which can affect the experience for many users based on factors such as ability (Dombrowski, Smith, Manero, & Sparkman, 2019) and gender (Makransky, Wismer, & Mayer, 2019; Maloney, Freeman, & Robb, 2021).

The need to see the product from a user's perspective resulted in various approaches to not only "consider" users but also directly involve them in the design process. Generally referred to as Participatory Design (PD), this extension to UCD tries to give a more active role to users (Abras, Maloney-Krichmar, Preece, & others, 2004). Various approaches to PD exist, such as co-design workshops (Eglash, et al., 2020; Lachney, et al., 2021; Nakamura, 2020), bringing experts from other disciplines into design group (Dopp, Parisi, Munson, & Lyon, 2019), and living labs as real-life environments for user collaboration (Dell'Era & Landoni, 2014; Pitse-Boshomane, et al., 2008; Veeckman, Schuurman, Leminen, & Westerlund, 2013). Based on one of the earlier definitions by Bergvall-Kåreborn and Stahlbrost (2009), Living Lab (LL) is a user-centric environment based on everyday practice and research to facilitate user participation in real-life contexts in order to create sustainable values. The five key components and five principles for a Living Lab approach are ICT & Infrastructure; Management; Partners & Users; Research; Approach: 1. Openness; 2. Influence; 3. Realism; 4. Value; 5. Sustainability (Bergvall-Kareborn & Stahlbrost, 2009). It is believed by LL proponents that the users must be understood in the context of their lived everyday reality, and the users must be involved as the co-creators in the design and innovation process (Ntšekhe, 2018). Ntsekhe also described how the LL environment allowed for utilizing multiple methods that help researchers to create a suitable solution.

The LL approach has been used and investigated by many researchers, especially in Europe, who have proposed different processes for implementing it (Schuurman, De Marez, & Ballon, 2016; Pierson & Lievens, 2005). However, the research on how to involve different user groups and communities in the research and design process is still in the early stages. This is particularly the case when dealing with emerging technologies where the community has limited knowledge and application areas, risks, and impacts are not clear.

### 3.3 Community-based Design

While successful in allowing designers to see the products from the user perspective, approaches such as UCD are limited as they base their process primarily on individual users. Groups of people interacting with each other introduce new dynamics that need to be considered in the design process. Fleury and Chaniaud (2023) discuss such an effect and suggest the notion of Multi-User-Centered Design. While important to establish the effect of user interactions in groups, such multi-user approaches to design are only halfway to a community-based approach. A community is more than multiple members and can be defined as a new holistic entity with its own traditions, protocols, and knowledge (Cattaneo, Giorgi, Herrera, & del Socorro Aceves Tarango, 2022; Guerrero Millan, 2023). Participatory design with communities can also be particularly sensitive when dealing with underserved and marginalized groups, due to historical context, access issues, existing perceptions, and unintentional harm (Harrington, Erete, & Piper, 2019). Participatory design workshops with such groups have the potential to be empowering if recognizing their situated nature and the sensitivities of the participants (Harrington, Borgos-Rodriguez, & Piper, 2019).

Preece et al. (2003) proposed Participatory Community-Centred Design, a process used primarily for developing online communities rather than engaging existing online communities. Lachney et al. (Lachney, et al., 2021) introduced the notion of community participation in Culturally Responsive Computing (CRC) with a focus on education. They invited local knowledge experts from different localized cultural systems, such as professional cosmetologists, farmers, and librarians, to join a design workshop with high school students. This resulted in an interdisciplinary collaboration that not only taught the students scientific knowledge from a textbook, but they also learned from the experts who have complementary tacit knowledge the students could further benefit from. Similarly, other researchers have suggested that more research on Indigenous communities should be conducted to bring out the perspectives of those with lived experiences, which will further help researchers to reveal the concealed histories by the mainstream media (GIDA, 2019).

Leong et. al. (2019) have concluded that, due to the heterogeneity among different indigenous communities, their needs, sociocultural practices, and ways to establish trust could be vastly different. Therefore, it is important to plan the research workshop with a lead user from different communities and obtain knowledge about each community early on, such as their daily routines and their calendars for community-related activities. Dutta (2019) has also suggested that research with indigenous communities should bring out the perspectives of the "insiders". O'Donnell et al. (2016) goes further to define the method of this community involvement by proposing a "whole community" approach with three levels of factors that shape digital technology adoption: (1) community members/household factors (mapping to the traditional TAM model), (2) community services and community-organization-level factors, and (3) local and transport infrastructure supporting individual and community adoption.

The International Indigenous Design Charter (Kennedy, Kelly, Greenaway, & Martin, 2018) is an example of introducing specific "protocols for sharing Indigenous knowledge in professional design practice". These protocols include addressing ethical concerns, allowing Indigenous representatives to participate in leading the projects, using objectives that are determined by the community, and involving deep listening on the part of the designer. Lewis et al. (2020) proposes similar principles for AI systems with Indigenous Peoples' data. They highlight locality, relationality and reciprocity, responsibility, governance, ethical design, and data sovereignty. These principles aim at ensuring that local communities are involved in the collection, processing, and governance of their data. Similarly, Steen (2022) explores ethical frameworks from Indigenous cultures such as sub-Saharan Ubuntu and Indigenous Wisdom from the Americas, as alternative to Western (or neo-colonial) method of design and application of technology, particularly AI, and emphasized on the need for reflexivity and care. While aimed at AI or technology in general, these works are essential in establishing a basis for a need to define inclusive approaches for technology design with communities focused on specific technologies rather than generic models. VR, just like AI, has its own special features and needs inclusion guidelines that are based on those features.

## 4  Inclusion in VR

The term Virtual Reality (VR) was originally used to refer to fully immersive experiences with Head-Mounted Display (HMD) devices but now refers to almost any Three-Dimensional Virtual Environment (3DVE), such as Immersive VR (IVR) with HMD (even though other VR platforms can also be immersive) or Desktop or Mobile VR (DVR and MVR, respectively). VR systems offer unique features such as immersion, presence, and interactivity plus flexibility (e.g., not constrained to physical laws), embodiment, and multimodality (ability to combine different media, input methods, and usage contexts) (Steffen, Gaskin, Meservy, Jenkins, & Wolman, 2019).

Mütterlein and Hess (2017) discussed factors such as technical features, ease of use, and trust influencing the acceptance and use of VR technology. Their study introduces human factors (trust, usability, age, etc) to the VR design and adoption process. Following up and based on Gibson's seminal work on affordance (Gibson, 2014), Steffen et. al. (2019) created a generalized framework of affordances for VR/AR relative to physical reality. Based on technical features such as immersion, simulation, interaction, and presence, they defined four affordances for VR compared to physical reality: diminishing negative aspects, enhancing positive ones, recreating existing aspects, and creating new ones. This framework can be used as a guide for the design and adoption of VR technology, is based on human factors, and implies inclusion as affordances directly relate technical features to people who can use them.

Our initial analysis showed that 106 sources on inclusion in VR could be grouped into four main categories based on the main inclusion subject: 22 for physical/cognitive ability, 32 for gender (with a few including other forms of personal representation), 27 for age, and 26 for community (ethnicity and culture). We used these four groups as the main inclusion themes in our study and the basis for structuring the inclusion discussion (Section 4). Some of our sources were related to more than one group (particularly, those discussing character representation), but we counted them towards the one that was the primary focus of the work.

Since the early years of VR, some researchers have raised awareness of inclusion by pointing out the individual differences and biases in character representation (Ford, 2001; Pate, 2020). With the recent development in affordable VR hardware, we have seen a huge spike in VR-related software tools, applications, and research studies. Even though the body of literature on VR has increased exponentially, inclusiveness is still lacking and is considered one of the biggest issues in the research community (Dombrowski, Smith, Manero, & Sparkman, 2019).

Through our review of the body of research on inclusion in VR, we identified two types of themes: those related to the user groups, and the ones related to inclusion concerns, as shown in Table 1.

A person can have various biological, social, and political characteristics, which may result in differences in how they access technology and participate in the design process. Our initial thematic analysis of the reviewed sources showed that the research on inclusion in VR had been mostly focused on the following themes:

- Accessibility based on physical and cognitive abilities,
- Special concerns related to gender,
- Working with children and the elderly, as specific age groups

- Cultural and ethnic considerations, particularly working with ethnic communities as opposed to individuals.

These were identified as the User Groups type of themes and were the basis of our general categorization. The following sub-sections (4.1 to 4.4) will review the work in these categories. It should be noted that a person's characteristics, such as age, gender, and race, are commonly intertwined and together form different states of discrimination and privilege, as described by intersectionality theory (Crenshaw, 2017). A proper understanding of inclusion in VR (and other areas) requires a more combined and holistic approach that should be the topic of future initiatives and seems missing in existing research.

Our thematic analysis also showed that the inclusion concerns for users were mainly about usability, representation, special concerns such as physical and cultural risks, customized applications, and the ability to own and develop their own VR experiences. We discuss these inclusion concerns, when applicable, within the next four sub-sections for each of the inclusion-related groups.

| Type of Theme | Description of Theme |
|---|---|
| User Group | Physical and cognitive ability |
|  | Gender |
|  | Age |
|  | Community |
| Inclusion Concern | Usability |
|  | Representation |
|  | Customized applications |
|  | Special concerns such as risks |
|  | Ability to own and develop |

Table 1. Inclusive Themes in Existing Literature

## 4.1 Accessibility in VR

Special-purpose VR applications have been used to help people with different mental and physical challenges and in various cases such as treatment and rehabilitation (Wilson, Foreman, & Stanton, 1997; Burdea, 2003; Montoya-Rodríguez, et al., 2022; Brepohl & Leite, 2023), learning and practice (Lannen, Brown, & Powell, 2002; Parsons & Mitchell, 2002), and general assistive purposes (Lagos Rodríguez, García, Loureiro, & García, 2022). While these special-purpose applications have been, at least potentially, successful in helping users, the research (Fernandes & Werner, 2022; Dombrowski, Smith, Manero, & Sparkman, 2019) suggests that the VR technology itself has accessibility issues such as lack of standards, reliance on limited modalities, and difficulties with walking and other forms movement. In this context, accessibility is defined as the design feature that makes the product usable by people with disabilities (Fernandes & Werner, 2022). The ability to customize different aspects of the VR experience is one of the main accessibility challenges (Phillips, 2020), and involving persons with disabilities in the design and development process, as opposed to an afterthought, has been suggested as an essential step to address these challenges (Mott, et al., 2019) (Phillips, 2020). Researchers have identified content, interaction, devices, representations, and applications as major accessibility opportunities for VR (Menke, Beckmann, & Weber, 2019; Mott, et al., 2019) and have questioned how VR can incorporate flexible training; how content can be presented; and how VR learning can be personalized.

A growing effort has been dedicated to making general-purpose VR systems more accessible. Dombrowski et al. (2019) follows the principles of Universal Design (Menke, Beckmann, & Weber, 2019; Connell, et al., 1997) and proposes seven pillars of accessible VR for equitable use, flexibility, simplicity and intuitiveness, perceptible information, tolerance for error, low physical effort, and size and space for use. The study also provided a 6-step plan for designing an inclusive VR game, which guides the game developer to be as inclusive as possible.

To address the over-reliance on vision, recent studies have explored alternative sensory experiences such as haptic (Siu, et al., 2020; Maidenbaum & Amedi, 2015) and have suggested tools such as magnification, text augmentation, and brightness/colour control (Zhao, et al., 2019). Researchers have also tackled the problem of limited movement for wheelchair users. Gerling et al. (Gerling, et al., 2020) employed techniques such as limited movement need, automated movement, and eye-tracking to control game avatars for wheelchair users. Qorbani et al. (2022) has noted the difficulties of wheelchair users with vertical and horizontal movement, especially within the small "safe" area of HMD devices. They explore the effectiveness of software controls and customizable virtual environments and show the potential for improving the accessibility of VR experiences in the case of a science lab.

To address accessibility as a high-priority issue, HMD manufacturers are beginning to include certain criteria for developers to embed accessibility features into their VR applications (Meta, 2020). However, there are limited academic studies on the application and efficacy of these features due to their novelty. Despite these efforts, the wide range of disabilities and possible solutions require more systematic studies. Multi-modalities, particularly the use of eye-tracking, various alternatives, customizability both in terms of environment (location of objects, colours, etc.) and controls, and the use of intelligent systems to predict and automate actions are among the most promising solutions.

While accessibility is the major usability issue, new research shows that people with disabilities share concerns on digital representation with many other ethnic and gender-based groups (Zhang, Deldari, Lu, Yao, & Zhao, 2022; Zhang, Deldari, Yao, & Zhao, 2023) that can affect usability. This emerging research emphasizes the importance of understanding the experiences of people with disabilities in social VR, especially in the context of avatar representation and interaction. Examples include the need to customize virtual canes and wheelchairs or creating safe space due to increased susceptibility to harassment and risks. Customized and specialized applications for people with disabilities, as opposed to generic ones that are accessible, can be another area of VR research that is less investigated.

## 4.2 Gender Bias in VR

Among the themes discovered in our analysis (Table 1), usability and representation are the most important ones when it comes to gender-based user groups, which show themselves as gender bias, i.e., the non-equitable status when using VR tecnologies. As for any other technology, gender-based differences in VR can be investigated in two ways: the potential (interests and capabilities) for use versus the actual use (existing targeted applications). Sieß et al. (2017) compared the changes in female users' interest in various aspects of VR and suggested that over the span of 25 years, this interest has grown significantly to levels like male users. Considering other studies, though, it is not conclusively shown that the interest and motivation in using VR are the same among men and women (López-Belmonte, Moreno-Guerrero, Marín-Marín, & Lampropoulos, 2022; Dirin, Alamäki, & Suomala, 2019).

The continuing difference in interest between genders regarding VR requires further investigation to verify because existing studies are not conclusive. It can also be related to perceived difficulties in using VR, lack of customized features, and gender bias in the design of VR systems. Munafo et al. (2017), in another early study about gender bias in modern VR technology, stated that the Oculus Rift HMD causes motion sickness and is sexist based on its effects. Their study showed that female users are more likely to suffer from motion sickness and other discomfort issues when using VR headsets, although the underlying reasons are still being investigated. A significant number of studies have confirmed this and suggested that other examples of gender bias exist regarding the headsets (more appropriate for men) (Lee O. , 2022), the sense of presence and perceived realism (higher for men) (Felnhofer, Kothgassner, Beutl, Hlavacs, & Kryspin-Exner, 2012; Melo, Vasconcelos-Raposo, & Bessa, 2018), perception of avatar (Schwind, et al., 2017; Regal, et al., 2022), learning and performance patterns (Makransky, Wismer, & Mayer, 2019; Grantcharov, Bardram, Funch-Jensen, & Rosenberg, 2003; Sagnier, Loup-Escande, & Valléry, 2020; Xia, Henry, Queiroz, Westland, & Yu, 2022), cybersickness (higher in women) (Shafer, Carbonara, & Korpi, 2017), and attention patterns (Porras-Garcia, et al., 2019).

However, these results have not been confirmed by other studies. For example, Melo et al. (2018) and Dayarathna et al. (2020), for example, finds no gender bias with respect to cybersickness and performance. These contradicting results may arise from several factors, including underrepresentation of female participants and authors in VR research, sample size, and test environments (Lopez, et al., 2019), which suggests that more comprehensive studies and more inclusive processes are required. But the findings are enough to encourage a more gender-sensitive approach to the design of VR experiences, both hardware and software (Paletta, et al., 2022).

The issues are even less investigated for LGBTQ+ VR users, who have received much less attention in research and development. The primary topic of research in this area is related to using VR as a tool for the LGBTQ+ community. Hernandez (2018) uses VR to fight for social justice, especially for LGBTQ+, "by acting as a digital archive of the experiences of sexually heterodox and gender variant migrant farmworkers in agricultural California." Acena and Freeman (2021) explore social VR, embodiment, and friendly interaction with diverse users as means of providing a space free of harassment, where they can comfortably communicate with others and present themselves. These examples exist at the limits of using existing VR tools but suggest many potentials for future work, such as the need for specific tools for avatar customization and the general digital representation of gender (and other aspects of) identity (Freeman, Maloney, Acena, & Barwulor, 2022; Freeman & Maloney, 2021; Morris, Rosner, Nurius, & Dolev, 2023), empathy for marginalized groups such as women and LGBTQ+ (Lee O. , 2022; Lui, Stringer, & Jouriles, 2021), safe interaction, intimacy, and social support (Li, Freeman, Schulenberg, & Acena, 2023; Freeman, Zamanifard, Maloney, & Acena, 2022).

Paré et al. (2019) begins to move beyond using VR as a tool and proposes the notion of "queering virtual reality," i.e., changing the design of VR systems to be more suitable for LGBTQ+ users.

Starting with Butler's heterosexual matrix (Butler, 2006), they question the "naturalness" of male/female binary categories. Then, they build on theories such as the feminist notion of the body becoming, i.e., considering the body not as a static entity but as a dynamic one that is constantly evolving, to suggest the need for moving away from presenting the body as constraining, fixed, and given towards a more dynamic and transformative process. This view can potentially change the avatar-building process in VR to accommodate better representation for gender non-conforming users. Proposing a more playful and intimate form of interaction, the authors discuss these open research questions: "1) specific forms of social support (e.g., informational, tangible, and emotional) that LGBTQ+ users could receive through social VR; 2) the unique role of social VR in LGBTQ+ users' construction and perceptions of their self-presentation online; 3) risks and challenges (e.g., privacy and harassment) emerging in LGBTQ+ users' engagement in social VR; and 4) how social VR can be designed to further support LGBTQ+ individuals' unique social needs." Focusing on the social role of VR, Maloney, et al. (2021) discusses topics such as identity and activity and raises questions such as: What are the considerations for designing a social environment inclusively for marginalized groups of all kinds? How can bias (e.g., gender/racial), body dysmorphia, and other concerns of self-identity be mitigated in social VR? How do we help protect personal space and other considerations of psychological and physical vulnerability in social VR?

### 4.3  VR for Elderly and Children

While age is continuous, and individuals at any age and at each stage of life can have their own specific conditions and concerns, most age-related studies in VR have focused on the elderly and children as two general target groups due to some common characteristics, and primarily identified usability issues and special concerns and risks for these groups.

Lee et al. (2019) developed a framework to evaluate factors such as the physical, social, and psychological well-being of AR/VR applications for older adults. According to their findings, a contextual framework and an evaluation framework for the critical review of the AR/VR technologies for promoting well-being in older adults were presented.

Soltani (2019) presented a SWOT analysis of VR for seniors that was created to understand the challenges the elderly is facing when accomplishing a mission and how VR can help them in overcoming these hurdles. It also mentioned the common characteristics of the system, the immersion, interaction, and stimulation. Soltani believes that VR has the potential to be used as an ecologically valid e-health screening system, which works best in addition to traditional methods (Soltani, 2019).

A growing number of studies have aimed at using VR by the elderly in various applications. VR is used to simulate architectural and environmental design for the elderly, such as service stations and retirement homes, for the purpose of evaluation (Wang Y. , 2022) or pre-occupancy training (Tseng, 2022). It is also used for rehabilitation (Qu, Cui, Guo, Ren, & Bu, 2022), physical/cognitive activity (Mortazavi, 2022; Dinet & Nouchi, 2022; Martínez, et al., 2022; de Farias, Montevecchi, Bokehi, Santana, & Muchaluat-Saade, 2022), and health training (Chang, et al., 2022).

While the results show potential in using VR, they are not conclusive and require further, more comprehensive studies (Martínez, et al., 2022). Also, most of these studies are aimed at having a more diverse set of applications for VR and less on making the technology itself more inclusive. Vahle and Tomasik (2022) investigated the effect of embodying an older avatar on social motivations. Their results suggest that visual cues such as avatars can cause social motivations commonly associated with old age. This had been observed in older studies where avatar appearance could influence behaviour and decisions (Peck, Seinfeld, Aglioti, & Slater, 2013). Arlati et al. (2022) explored the differences in kinematics and showed that "older adults moved slower, more curved, and reached lower peak velocity" compared to younger VR users. They hypothesized that these differences might be related to vision and cognitive ability, VR experience, or haptic feedback but concluded that further investigation is needed to understand how the elderly use VR and how VR experiences need to be designed to match their use.

Similarly, VR use by children is drawing the attention of many researchers. Fonseca et al. (2018) conducted a mixed-method study to assess how wearable VR can affect usability and satisfaction, as well as how VR technologies can improve the learnability of students with ADHD. Even though this project is still in its early stage, the quantitative and qualitative methods employed in this study have proven useful as its educational objectives. However, the results showed a lack of usability of the system, mainly due to the mobile UI design and the network issues between students' devices and multimedia content. Though, the authors claim that this system provided a better understanding of the museum's characteristics to the students.

Vasconcelos et al. (2017) created a protocol for creating serious VR games to aid intellectually disabled children to facilitate alphabetization and the teachers. The proposed and prototyped VR-based game carried out several learning objectives for intellectually disabled children by simulating day-to-day activities such as object seeking, information seeking and information comprehension. After conducting user studies and evaluating the study results, educators validated the protocol, and the serious game can be used as a tool to support education.

More recent studies have used VR, particularly VR games, to help children with visual attention (Flores-Gallegos, Rodríguez-Leis, & Fernández, 2022), social skills (Ke, Moon, & Sokolikj, 2022), motor skills (Jung, Chang, Jo, & Kim, 2022), cognitive deficit (Corrigan, Păsărelu, & Voinescu, 2023), empathy development (Muravevskaia & Gardner-McCune, 2023), and fear and anxiety (Hsu, et al., 2022; Sülter, Ketelaar, & Lange, 2022), among other objectives. Like the VR applications for the elderly, these studies generally show the potential of VR but demonstrate multiple issues, mainly inconclusive results that can be attributed to the quality of studies and VR experiences, small sample size, and short evaluation periods. This suggests that further, more comprehensive studies are required. At the same time, researchers have started to point out possible risks that VR can pose to children (Kaimara, Oikonomou, & Deliyannis, 2022). Overall, it seems that further research is needed to understand how VR experiences should be designed for children and what specific risk factors exist. Existing literature on design for children (Hourcade, 2008) should be considered and applied properly to VR experiences. The role of children as active participants in the design process is also an important issue which is not considered properly in the design of VR experiences for children (Goagoses, et al., 2022).

## 4.4 Using VR for Communities

VR has been used extensively to represent different cultural and community features such as language, art, history, and lifestyle, in addition to limited research on the use of VR for marginalized groups, with empathy and shared experience among the most common goals, especially for black communities (Peck, Seinfeld, Aglioti, & Slater, 2013) (Thériault, Olson, Krol, & Raz, 2021). As the point of this review is the notion of inclusion in VR, a full review of these VR applications is beyond our scope. But to better understand the positioning of VR in ethnic and cultural communities, we offer a brief set of examples. We then look at some approaches to the design process of VR experiences with and about these communities and finally review the inclusion-related concerns for the specific case of Indigenous groups as a salient example. Among the inclusion concerned that we identified in our analysis (Table 1), representation, customization, ownership, and ability to develop were highlighted for community-based cases.

### 4.4.1 Examples of using VR for Community-related Applications

Chen et al. (2010) emphasized the importance of consistency and harmony in digital heritage, suggesting that edutainment is an effective technique for engaging users. Manzhong (2017) argues that using VR can greatly promote the sharing of traditional cultural resources in society and provide spiritual impetus and intellectual support for promoting advanced culture (2017). Schofield et al. (2018) created a VR experience with an interdisciplinary team to demonstrate Viking culture in a museum. The study focused on representing the Viking culture with authenticity, informativeness, and novelty of the VR experience. Even though the user study was preliminary, findings showed how the visitors focused more on the content rather than the technology. One of the technical challenges of this study was how to ensure prolonged and cost-effective technical support in such a museum environment. Moreover, one of the design challenges encountered by Schofield et al.'s interdisciplinary team was the difficulties in developing and assembling the scenes together.

Fu et al. (2020) proposed a design method that retains the key form of professional knowledge while reducing users' learning barriers through interactive experiences. Gao et al. (2021) conducted a study with 70 Chinese university students, comparing the effectiveness of traditional classroom instruction with virtual reality-based instruction in enhancing intercultural sensitivity. Du (2022) posited that VR technology could be utilized to create three-dimensional, interactive, and immersive experiences, allowing users to better understand and appreciate the richness and complexity of intangible cultural heritage.

Li (2022) identified five core concepts that influence the digital dissemination of intangible cultural heritage (ICH) handicraft using VR: digital communication awareness, cultural communication initiative, adaptability of digital technology, audience acceptance and cognition, and uncontrollable factors. Liu (2015) designed a distributed, scalable, interoperable service platform to enable unified retrieval and revelation of digital cultural resources. The system is open to six cultural institutions and includes ten million metadata and one million object data with various media types. Wang et al. (2022) explored the relationships among virtual reality tourism involvement (VRTI), place attachment, and behavioural intentions, providing further insight into the potential of VR technology in cultural experiences.

Overall, these studies demonstrate the versatility of VR in promoting cultural engagement and education while preserving the details and essence of cultural heritage. By using a variety of technologies and methods, VR can allow users to explore cultural heritage in new and exciting ways while also preserving these important sites and artifacts for future generations.

### 4.4.2 Design Process

The domain of the multi-user VR environment brings inevitable biases during the design and development process and the product (Ford, 2001). Ford (2001) suggests that designers and researchers should pay attention to their users' demographic as they make decisions on avatar representations and the user choices of how to be represented. Moreover, the use of NPC (bots) in VR should consider the implication of the 3D models that are used on the bots. For example, a shoe-shining bot that looks like a young African American male would cause social implications (Ford, 2001).

VR researchers and developers have recently started to consider participatory design methods (as discussed in Sections 3.2. and 3.3) more actively in their work. By reviewing the existing works, we notice that beyond technical aspects such as realism of rendering or performance-efficient 3D modelling, the design approach is primarily about authenticity and other considerations (such as ownership) from the people of the related communities. For example, Loudon et al. (2019) uses design thinking sessions to include Indigenous communities of India in the technology development process. Marques et al. (2021), similarly uses participatory design when working with communities in Australia and New Zealand.

The use of Participatory Design (PD) methods has been widely recognized as an effective approach in the development process of Virtual Reality (VR) applications (Kong & Zhang, 2021; Park N. , et al., 2022). The primary objective of employing PD methods in this context is to explore the potential benefits of facilitating knowledge transfer from workshops to the demographic involved (Steen, 2022; Leong, Lawrence, & Wadley, 2019; Wallis & Ross, 2021). While surveys and questionnaires are commonly used as the primary means of conducting user studies, literature has highlighted the effectiveness of PD methods, such as co-creation and co-design, in enhancing engagement and knowledge transfer between traditional handicraft practitioners and users (Wallis & Ross, 2021). These methods can also enable the development of VR experiences that are more relevant and accessible to the target demographic.

As a subject example, China is one of the most populous ancient countries, which has a long history of cultural evolutions where the activities are vastly different from location to location. The country may have only one official language, but the context and nuances of the cultural activities could get lost in translation. So, the involvement of those who are being represented becomes critical in the process. Fu et al. (2020) conducted a study in which they compared the VR experience with the regular method of disseminating cultural heritage. The study involved working with both experts and regular users and providing an environment for users to conduct activities within the VR environment. The study yielded positive feedback and promising outcomes. Although the experts were not actively involved in the design process, the study highlights the efficacy of employing PD methods to enhance knowledge transfer and engagement in VR development processes. Kong and Zhang (2021) proposed a community-based participatory design model that involves organizers considering the interaction protocols of participants and applying appropriate PD methods to local cultural backgrounds. The model comprises three stages, wherein the role positioning and organizational relationships among participants change throughout each stage.

Overall, researchers are realizing that the relationship of communities with technologies such as VR is more complex than what is described in existing models; their adoption/approval of a technology depends on many community structural elements and requires their involvement in the design process (O'Donnell, et al., 2016; Yong, Arya, & Manatch, 2023; Kong & Zhang, 2021). Participatory design process, as such, need to be customized to match community needs and values (Yong, Arya, & Manatch, 2023). Considering the significant focus of existing research on Indigenous Peoples as examples of communities with clear and distinguished traditions, in the next section we will revie some of the approaches for such participatory design.

### 4.4.3 VR and Indigenous Communities

### 4.4.3.1 Indigenous Communities and Digital Technologies

Beyond the realm of VR, researchers have paid increasing attention to proper ways of working with Indigenous communities, with a new understanding that such work should be "by" or "with" as opposed to only "for" or "about" the community. It should go beyond simple participation to empower communities to develop and own the end products based on their traditions and protocols (Peters, et al., 2018). Various forms of co-design and participatory research have been studied. Park et al. (2022) investigate the challenges of the Maori of New Zealand in preserving their cultural heritage. A co-design process is used, with mixed-reality as a medium, to help community engage in their culture and language. The authors emphasize on the importance of respecting and engaging the community in the design process. Taylor et al. (2022) proposes the notion of tangible design as a form of community-based co-design within the context of working with Australian Aboriginal and Torres Strait Islander communities. Mukumbira and Winschiers (2023) emphasize the role of epistemological differences and how Indigenous worldviews and protocols should be considered in the design of digital technologies for Indigenous communities. This is in line with the notion of ontological and pluralistic design (Escobar, 2018) and reinforces the involvement of Indigenous people in research and development projects about them to ensure proper representation (Iseke-Barnes & Danard, 2007).

Campbell-Meier et al. (2020) has summarized four digital inclusion elements studied in the literature, Access, Skills, Motivation and Trust:

- **Access:** Young (2019) conducted a case study with an Inuit community in the Canadian Arctic. The methods involved were participant observation, archival research, and semi-structured interviews. Young suggested that blindly providing digital devices to the indigenous communities may also cause negative effects which may endanger the communities' traditions and culture and found that digital device erodes critical components of Inuit knowledge. According to the participants of Young's study, most of the community elders have pointed out that the younger generation has spent more and more time on their digital devices and is less motivated to participate in the learning activities organized by the local community. In a contrasting or complementary study, Rice et al. (2016) believes that using the digital device could enhance stronger cultural indignity and community and family connections. They conclude that indigenous young people use social media to help form, affirm and strengthen identity, feel a sense of power and control over their own lives, and make and continue community and family connections. Moreover, these young people also used social media to transmit intergenerational knowledge, improving persona, family and community relations between the young and old generations. However, Rice et al. have also pointed out the negative aspects of using social media for young indigenous people, which include cyberbullying, racism, and sexting.
- **Skills:** Based on Campbell-Meier et al.'s review (Campbell-Meier, Sylvester, & Goulding, 2020), skills represented the smallest proportion of the body of literature reviewed. The authors believe that digital literacy skills were often not the focus of the research checked, but skill development has acquired more attention. This is later shown to be an important issue when it comes to emerging technologies like VR (Yong, Arya, & Manatch, 2023). Educators around the globe have teamed up with Indigenous communities to promote Indigenous cultures but also expose Indigenous communities to modern technologies. Stanton et al. (2019) conducted a case study that evaluates the Digital Storywork Partnership (DSP) to advance the goals of Indian Education For All (IEFA, https://opi.mt.gov/Educators/Teaching-Learning/Indian-Education-for-All), which aims to bring the traditional classroom through community-centred research and filmmaking. The study showed that DSP is culturally affirming and revitalizing for Indigenous communities and holds the potential for use in all cities. While this study did not introduce anything regarding ICTs, the collaboration between the educators and the Indigenous population has shed light on a possible way to teach

- Indigenous cultural lessons to the participants. During the filmmaking process, the community members are all involved, strengthening the quality of learning for the younger generation. Stanton et al. suggested that this process has helped young Indigenous students to develop a complex awareness of their community.
- **Motivation:** McGinnis et al. (2020) tried to incorporate the interview data into digital tourism products to promote the local indigenous communities and evaluated if the community-based technological development could motivate the community members with better effectiveness. In another user evaluation study (Shiri & Stobbs, 2018) about an Indigenous cultural heritage digital library in northern Canada, not providing multilingual support during the early phases of the study negatively affected the motivation of the community members.
- **Trust:** Maitra (2020) suggested that humans can mitigate the sense of "loss of control" to artificial intelligence by borrowing and learning the "non-human" aspect of the relational nature of the indigenous philosophical and practical interconnection and interrelatedness beyond the Eurocentric "flourishing of human. So, human and AI could find a way to "co-exist" in the future. Maitra's paper is an atypical work from our field. It discussed how Indigenous values and cultures could create a particular relational schema or non-human soul-bearer from the indigenous epistemology. Maitra discussed AI from indigenous perspectives on human rights, value alignment, ethics, and how our attitudes could shift towards AI. This idea looked at the spiritual beliefs of the indigenous culture. Maitra believes that for humans and AI to co-exist in the future, researchers could seek assistance in Indigenous traditions (Maitra, 2020).

Inspired by Indigenous research methodologies, McMahon's study (2020) paid a fair bit of attention to the policy side of digital inclusion with indigenous communities, which raised several points on how creating "demand-side interventions could support digital inclusion and adoption". McMahon worked with indigenous partners to create digital literacy learning courses as a multi-day training program for secondary schoolers. The process of preparing and executing this program brought many positive outcomes. For example, both the elderly and the youth actively participated in different aspects of the camp and had positive experiences all around. Providing an actionable plan seems a useful "demand-side" intervention to support digital inclusion. Lastly, McMahon also suggested six areas for future digital inclusion initiatives. While the bulk of them is policy-related, the last site presented is "developing appropriate digital literacy resources," which aligns with the digital literacy program they have developed for the indigenous community.

### 4.4.3.2 Indigenous Communities with Inclusive VR

Virtual reality applications have been used frequently to represent different aspects of Indigenous culture and offer educational possibilities. In particular, the Indigenous language (Kelly & Russell, 2023), traditional environmental knowledge (Dawson, Levy, & Lyons, 2011), and storytelling (Arendttorp, et al., 2023) have been the subject of many Indigenous VR experiences.

Fewer VR studies have aimed at offering customized services for Indigenous communities, with different levels of community participation in the design process. Seon et al. (2023) reports on VR-assisted Cognitive Behavioural Therapy for Inuit communities of Canada. They emphasize the role of community involvement to make sure the products are suitable for community members and based on their specific needs and concerns. Regenbrecht et al. (2022) defines the principles of partnership, participation, and protection within the context of a telepresence VR experience for Māori community of New Zealand. They describe their collaborative process and take steps toward a more formal community-based design process that can be used for other cultures (Regenbrecht, et al., 2022) (Park N. , et al., 2022).

The advances in VR hardware and software, combined with the increased availability, have made VR an attractive solution for creating engaging experiences related to Indigenous storytelling and education (Wyeld, et al., 2007). While such VR applications for and by Indigenous Peoples are increasing, important questions have remained unanswered, such as how to ensure authenticity and ownership. Rodil (2015) argues that using Information Technology by the community means dealing with a foreign design, which suggests the importance of participatory design processes.

In a recent study, Wallis and Ross (2021) discuss an Indigenous-centred VR production framework within the context of sample projects. They show that existing productions present different levels of involvement from Indigenous Peoples (from user to participating or main designer) and repeated trends such as the use of "VR to express and realize Indigenous futurism; provide new articulations of Indigenous activism; and embody connections." Leong et al. (2019) shares their insights regarding PD with the Indigenous Peoples of Australia. They ran a series of workshops with Indigenous communities on the use of mobile devices and social media to connect those communities and the rest of the country. They suggested guidelines such as recognizing the differences between various communities, establishing trust between different Indigenous communities, having community-led workshops, establishing connections with the Indigenous communities before the workshops, and respecting the Indigenous calendars to avoid conflict with the pre-existed Indigenous activities.

Wallis and Ross (2021) helped establish an indigenous-centred VR production framework while working with the community to understand how VR is being used to create space and capacity for indigenous creatives to tell their stories and how indigenous creatives negotiate Eurocentric modes of production and distribution. The community can produce VR content to express and realize indigenous futurism; foreground native languages in virtual worlds; provide new articulations of indigenous activism; embody connections between the past, present, and future and demonstrate the interconnectivity of all living things. Like Radianti et al.'s work (Radianti, Majchrzak, Fromm, & Wohlgenannt, 2020), the knowledge acquired through digital transmission to the local community is seldom studied.

The limited number of participatory VR design projects does not allow a proper definition of a design process customized for VR experiences with and for Indigenous Peoples and other ethnic communities. In one of the latest works, Yong et al. (Yong, Arya, & Manatch, 2023) propose Indigenous Technology Empowerment Model (ITEM). Through the example of a VR-based Gathering Place for an Indigenous community in Canada, they build on the notion of empowering through active participation and ownership, as opposed to simply adopting technology (Seth, 2022) (Williams Goodrich, 2019). Yong et al. (2023) defines a framework that includes Technology, People, and Content on the foundation of Process, as a model for emerging technologies such as VR when used for/by/with Indigenous People. The model can be generalized to other communities and emphasizes the important role of Process with elements such as integrated community participation and education that empower the community to develop and own its VR experiences.

## 5    Discussion

Our review started by acknowledging that different individual and group characteristics must be directly considered in the design of VR experiences to ensure proper inclusion. Recognizing that there are many differences among users, where each and any combination can be the cause of discrimination and privilege, we focused on some that were the most highly discussed in the literature, i.e., ability, gender, and age, in addition to community-based inclusion with special attention to Indigenous Peoples as a clear and important example. We aimed to answer three main questions about explored barriers, knowledge gaps, and research directions. These questions are discussed in sub-sections 5.2 to 5.5 for the main four user groups (Table 1), respectively. But first, we discuss some general insights gained from our review.

## 5.1 General Insights

While different groups of users who were the topic of our review had their own specific concerns, our literature review revealed some general insights that almost equally apply to all inclusion-related research and design efforts. Some of the major examples of these insights are as follows:

- **Inclusion-related concerns can be grouped into typical categories.** As shown in Table 1, these categories include items such as usability, representation, risks, and ownership. More research is needed to investigate other concerns related to inclusion.
- **Intersectionality needs to be incorporated into inclusion research**. Almost all existing research studies consider user characteristics in isolation and do not investigate their intertwined nature.
- **Most inclusion-related studies are focused on ability, gender, and age**. In addition to a combined approach, the research studies should also pay attention to other factors that may be barriers to inclusion. The level of education, profession, lifestyle, cultural background, geographical location, personality type and traits, and many other personal factors can influence the experience of users with VR (and other technologies) and, as such, need to be investigated.
- **Inclusion-related studies are commonly performed on small sample groups, for a short duration of time, and using one or two example applications and experiences**. For example, a study on the effect of gender in VR-based learning should use a reasonably large sample size, be longitudinal, and include multiple subjects and scenarios in order to be generalizable.
- **The role of manufacturers, designers, and developers in inclusion is seriously under-studied**. Very few research studies have paid attention to what these groups do, say, and feel about incorporating inclusive aspects in their work, what difficulties they face to do so, and how efficient various tools and solutions may be.
- **The research on inclusive participatory design in VR is in its infancy**. This is particularly a problem because VR and other emerging technologies are complicated, and involving end-users who have no proper education is hard. Finding ways to educate and engage users in the design process is crucial in reaching inclusion in VR technology and applications.

## 5.2 Accessibility

Researchers have investigated the accessibility of VR experiences mainly in the context of content, interaction, and devices. It has been recognized that there are no clear accessibility standards in VR, which is a fundamental barrier for people with disabilities when using VR. Some new manufacturer guidelines and requirements are being introduced that can help with this problem, but they are not necessarily compatible and not based on more general ones like Universal Design. The lack of support for different modalities is another major accessibility barrier caused by overreliance on vision and mobility.

In addition to the barriers identified earlier, our review showed that there is a serious gap in research and knowledge on proper ways to design VR experiences for those with limited abilities, especially in movement and vision. Existing research has started to shed some light and propose methods, but it is not conclusive and requires further investigation. Similarly, there is limited research on special and customized applications for people with disabilities.

Moving forward, hardware and software customization and (personalized) alternatives seem to be a promising direction. Virtual environments can offer easy-to-use interface alternatives and audio/visual settings to accommodate the needs of different users. Another research direction should be towards a common implementation of standards and guidelines, such as Universal Design in the context of VR. Finally, more research studies are needed targeting both users and developers and a wider range of applications. Some specific research questions that need further investigation are:

1. How do we support and combine different modalities in VR?
2. What is a proper standard accessibility protocol for VR?
3. How can we use software customization to optimize a VR experience for people with disabilities?

### 5.3 Gender

It is suggested by many studies that VR hardware and software can have a gender bias against women and particularly LGBTQ+ users, that prevents them from benefiting equitably from the technology and application. Initial studies have proposed methods such as hardware and avatar customization and some safe space guidelines, but neither the study of gender bias nor the proposed solution are conclusive.

There is a clear research gap to (1) conclusively and through more comprehensive studies investigate gender bias and how it may affect the users and (2) find out what hardware and software methods can avoid and eliminate this bias.

Future research should explore the notion of customization beyond avatars and pay particular attention to interaction, communication, and representation methods. The notions of identity and safe space in virtual environments must be investigated, and proper ways of achieving them must be identified. Except for some work on increasing empathy, topics such as social support, intimacy, and perspective change in VR have received almost no attention in current research and are directly related to gender bias. Some specific research questions that need further investigation are:

1. What VR activities and features have a gender bias?
2. What are specific concerns for LGBTQ+ VR users?
3. How can we achieve better representations in VR beyond avatar customization?
4. What is a VR safe space, and how can we implement it?

### 5.4 Age

Current hardware and software are not optimized for the elderly and children due to multiple factors such as vision, cognitive ability, and familiarity. While there are relatively more VR studies targeting older people, due to various risk factors, using VR for children is still facing resistance, which, as shown by a few existing studies, may mean children are not benefiting from this potentially useful technology.

The barriers to properly implementing VR solutions for the elderly and children are mainly related to the lack of knowledge of their special needs and risks. These special concerns and the effectiveness of VR experiences for these target groups are severely under-studied. There are also clear gaps in how the general guidelines for design for children and the elderly should be customized for VR.

Future research should target the above research and knowledge gaps and actively include children in the design process. Some specific research questions that need further investigation are:

1. What are the specific risks in VR experiences for children and the elderly, and how can we manage them?
2. What are proper user interfaces and interaction styles for different age groups?
3. What are activities that are best suited to VR for children and the elderly?

## 5.5 Community

Researchers and practitioners are becoming increasingly aware that working with communities can not follow the same process as working with a group of diverse individuals. A community has its own tradition, knowledge, concerns, and protocols. Particularly for marginalized communities, it is important to include the community directly in the process and empower them to create and own technology products that benefit them. The primary barrier to such inclusion is that there is no clear model or set of guidelines for using VR with/for/by communities. The second important barrier is the lack of proper education that enables active participation. Larger groups, multiple applications, different communities, and longer evaluation time are needed to investigate both the community participation models and the integration of education.

Some specific research questions that need further investigation are:

1. What are the best practices of the design process for VR with communities, especially Indigenous Peoples?
   a. What specific concerns, interests, or protocols do these groups have for the use of VR? Examples include ownership, authenticity, respect, and co-design.
   b. What specific affordances can VR provide for these target groups? Examples include interactivity, engagement, immersion, and visualization.
   c. What technical features can be developed to address the identified items from questions a and b?
2. How can existing VR design and development methodologies be improved to be more inclusive?
   a. What structures can the design teams have to incorporate community members?
   b. How can we integrate training and education to allow more active participation?
   c. What are the roles the community members can play besides providing input early and doing the evaluation?
   d. How can we choose or mix the methods used when conducting studies with marginalized communities?

User-centred and participatory design methods (including the latest examples such as Living Labs) have been widely used with a wide range of demographics, but they have not been proven with the Indigenous and other under-represented communities. It is necessary to develop and refine these approaches to match the needs of community-based research and design. Participants selections, study design, the before-study preparation and after-study debriefing and follow-up, and research/design team structure are among the process aspects that can be customized.

While most of the above questions are applicable to other technologies, it is essential to try to answer them within the specialized context of VR and its unique features.

## 6  Conclusion

In this review, we aimed to identify the barriers, research gaps, and future research directions to increase inclusion in VR technology and experiences. We showed that the existing research has focused on ability, gender, and age as personal characteristics and distinguished between individual vs. community-based design, with a special focus on Indigenous communities. Our review identified the strengths and weaknesses of current research and suggested general and specific themes for future research.

In addition to the direction that we identified for future research, similar reviews are also needed to look in more detail into the design aspects of existing studies to identify their explicit and implicit inclusionary decisions, especially beyond ability, gender, and age. Also, VR work with communities other than Indigenous People (ethnic, cultural, and other forms of communities) should also be reviewed. A literature review focused on VR, and communities can be helpful in this regard. In the absence of any comprehensive review about Inclusion in VR, we hope that our review can act as the first step towards a general model of inclusion in VR technology and applications.